\documentclass[acmsmall]{acmart}
\usepackage{soul}
\usepackage{subfloat}
\usepackage{graphicx}
\usepackage{subfigure}

\soulregister\cite7 
\soulregister\citep7 
\soulregister\citet7 
\soulregister\ref7 
\soulregister\pageref7 

\AtBeginDocument{%
  \providecommand\BibTeX{{%
    \normalfont B\kern-0.5em{\scshape i\kern-0.25em b}\kern-0.8em\TeX}}}

\setcopyright{acmcopyright}
\acmJournal{TOMM}
\acmYear{2020} \acmVolume{1} \acmNumber{1} \acmArticle{1} \acmMonth{1} \acmPrice{15.00}\acmDOI{10.1145/3418211}

\begin{document}

\title{Market2Dish: Health-aware Food Recommendation}

\author{Wenjie Wang}
\email{wenjiewang96@gmail.com}
\affiliation{%
  \institution{National University of Singapore}
  \city{Singapore}
}

\author{Ling-Yu Duan}
\affiliation{%
  \institution{Peking University}
  \city{China}
  }
\email{lingyu@pku.edu.cn}

\author{Hao Jiang}
\affiliation{%
 \institution{Shandong University}
 \city{China}
}
\email{jianghaosdu@mail.sdu.edu.cn}

\author{Peiguang Jing}
\affiliation{%
  \institution{Tianjin University}
  \city{China}
}
\email{pgjing@tju.edu.cn}

\author{Xuemeng Song}
\affiliation{%
  \institution{Shandong Unviersity}
  \city{China}
}
\email{sxmustc@gmail.com}

\author{Liqiang Nie}
\affiliation{
  \institution{Shandong University}
  \city{China}
}
\email{nieliqiang@gmail.com}

\renewcommand{\shortauthors}{Wang and Duan, et al.}

\begin{abstract}
 With the rising incidence of some diseases, such as obesity and diabetes, the healthy diet is arousing increasing attention. However, most existing food-related research efforts focus on recipe retrieval, user preference-based food recommendation, cooking assistance or the nutrition and calorie estimation of dishes, ignoring the personalized health-aware food recommendation. Therefore, in this work, we present a personalized health-aware food recommendation scheme, namely Market2Dish, mapping the ingredients displayed in the market to the healthy dishes eaten at home. The proposed scheme comprises three components, namely recipe retrieval, user health profiling, and health-aware food recommendation. In particular, recipe retrieval aims to acquire the ingredients available to the users, and then retrieve recipe candidates from a large-scale recipe dataset. User health profiling is to characterize the health conditions of users by capturing the textual health-related information crawled from social networks. Specifically, to solve the issue that the health-related information is extremely sparse, we incorporate a word-class interaction mechanism into the proposed deep model to learn the fine-grained correlations between the textual tweets and pre-defined health concepts. For the health-aware food recommendation, we present a novel category-aware hierarchical memory network based recommender to learn the health-aware user-recipe interactions for better food recommendation. Moreover, extensive experiments demonstrate the effectiveness of the health-aware food recommendation scheme. 
\end{abstract}

\ccsdesc[500]{Information systems~Multimedia and multimodal retrieval}
\ccsdesc[500]{Information systems~Collaborative filtering}
\ccsdesc[500]{Information systems~Recommender systems}
\ccsdesc[500]{Applied computing~Health care information systems}

\keywords{User Health Profiling, Health-aware Food Recommendation, Recipe Retrieval.}

\maketitle

\section{Introduction}
Food is essential for human beings. Once people's basic requirement for food is satisfied, they focus on the pursuit of a healthier diet. Nowadays, countless people are being plagued by many diseases due to the unhealthy diet.
According to the World Health Report 2018 \footnote{\url{https://www.who.int/gho/publications/world_health_statistics/2018/en/}.}, the incidence rate of many diet-related diseases is increasing rapidly over the world, such as diabetes, obesity, and malnutrition. Based on the healthy diet tips\footnote{\url{https://www.meishij.net/}.}, people with different health conditions ought to be endowed with a personalized healthy diet. For example, the diabetics are required to eat more whole-grain cereals and avoid sweet food. However, people are usually caught in a dilemma when deciding what ingredients they should buy from the market according to their physical conditions. Many reasons lead to such dilemma: 1) People's personal knowledge and experience on food are limited, while a healthy and delicious dish is usually complex, involving many ingredients and cooking skills. And 2) many people cannot explicitly and precisely describe their own health conditions, let alone correctly judge what kinds of food are healthy ones.

Despite the great value of a health-aware personalized food recommendation system, there still exist many challenges: 1) The amount of ingredients available in the market is huge. It is non-trivial to learn a mapping function between the ingredients available to users and their expected food. 2) How to obtain the health profile of a given user needs study. After all, the health-related user information we can acquire from users or the Internet is extremely sparse.
And 3) jointly considering the existing food knowledge and the health profiles of different people to recommend the personalized healthy ingredients or food is a tough issue. In this work, we try to overcome the aforementioned challenges by proposing a health-aware food recommendation framework, which aims to profile user health, and then recommend the healthy food to users.

Actually, the rapid expansion of Internet has provided much information for us to solve the aforementioned challenges: 1) The popularity of smartphones enables us to capture and record our lives visually and vividly, from which rich ingredient information in the market can be obtained. 2) People are keen to enjoy social networks, such as Twitter\footnote{\url{https://twitter.com/}.} and Weibo\footnote{\url{https://www.weibo.com/}.}, and share personal information on these platforms, such as activities, likes, and dislikes. These social media platforms inadvertently expose personal health information more or less. For example, it can be easily found out that one is losing weight while the other might get pregnant from their shared tweets. And 3) the explosive growth of online information brings us massive food data as well. Many high-quality and food-related data is available in some recipe-sharing websites (\textit{e.g.},  Yummly\footnote{\url{http://www.yummly.com/}.} and Meishijie\footnote{\url{https://www.meishij.net/}.}). In addition, extensive food-related health knowledge can be acquired in many forms, such as what kind of food will make you fat or lower your blood pressure.

Indeed, several prior methods on food recommendation have paid attention to satisfy users' taste preferences by modeling the historical user-item interactions and predicting the future ones \cite{harvey2013you}. Nowadays, health-aware food recommendation has become an emerging research topic. For example, Ge \textit{et al.} \cite{ge2015health} leveraged the so-called ``calorie balance function'' to incorporate calorie counts into the food recommendation method. Elsweiler \textit{et al.} \cite{elsweiler2017exploiting} explored the feasibility of substituting meals recommended to users typically with similar but healthier dishes with the help of the user study.
Although great success has been achieved, they failed to build user health profile and recommend them personalized healthy diets based on the available ingredients.

%
\begin{figure}
\vspace{0cm}
\includegraphics[scale=0.6]{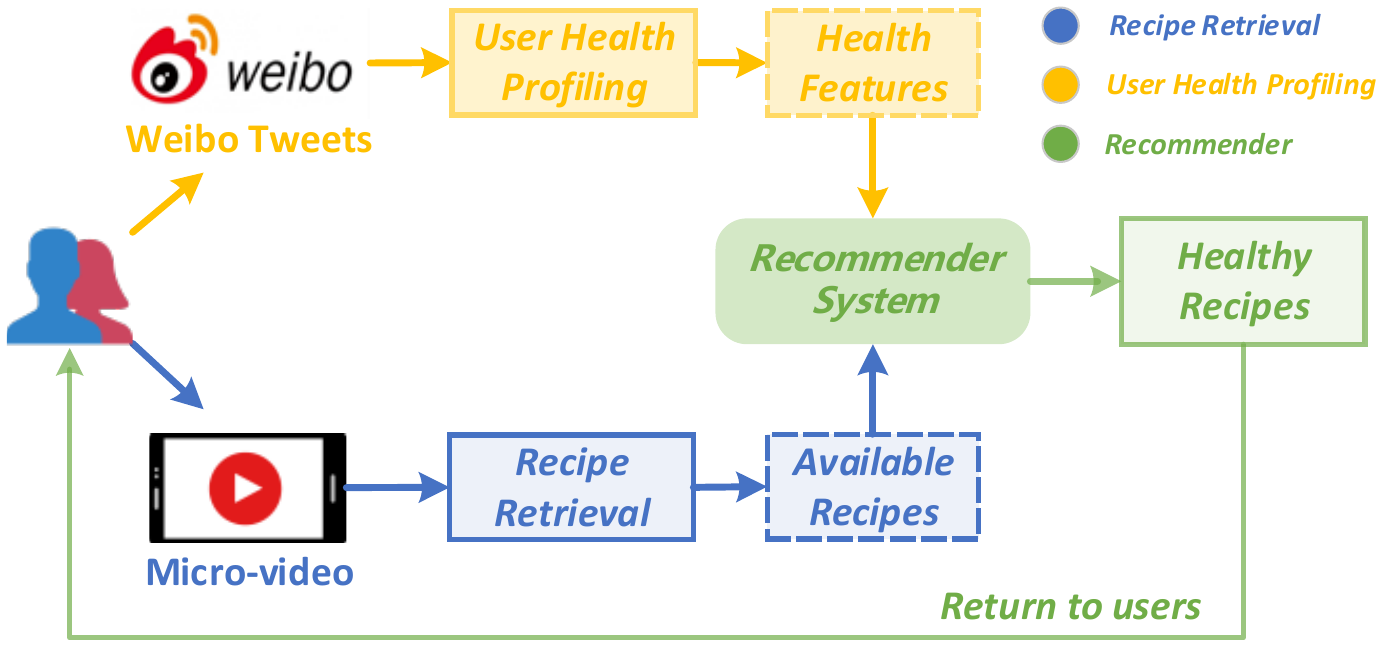}
\caption{Illustration of our proposed scheme. Recipe retrieval is applied to retrieving the recipe candidates for users; meanwhile user health profiling aims to extract users' health features; ultimately, the recommender returns the personalized healthy recipes to users.}
\label{Figure1}
\end{figure}
%
%

To help users decide the personalized healthy diet, we propose a health-aware food recommendation scheme, as illustrated in Figure \ref{Figure1}. It comprises three components: recipe retrieval, user health profiling, and recipe recommendation. To be more specific, 1) on recipe retrieval, users have many ways to input their available ingredients from markets, such as selecting from ingredient candidates, voice input, and micro-video input. In this work, we explore the possibility of micro-video input, by which users can record various ingredients in the market by a micro-video conveniently, and then Inception-v3 Net \cite{szegedy2016rethinking} is utilized to capture ingredients via the multi-label image classification. Based on the available ingredients, we can retrieve many recipes for each user from a large-scale recipe dataset. Due to the high accuracy of existing methods on image/video classification, we focus more on the following two components in this work. 2) As for user health profiling, we pre-define some health concepts to represent the common health conditions with the help of domain experts. The health concepts are collected from different perspectives, such as age  (\textit{e.g.}, teenagers and old people), occupation (\textit{e.g.}, office workers and students) and diseases (\textit{e.g.}, insomnia, hypertension, obesity, and malnutrition). And then user health profiling aligns users with the pre-defined health concepts by capturing the textual health-related information crawled from users' social accounts (\textit{e.g.}, Weibo). In this way, user health profiling is converted to a text classification problem. However, existing text classification methods usually learn a high-level latent representation for textual content, which is overwhelmed by much noise since the health-related information collected from social platforms is extremely sparse. To alleviate this issue, we present a word-class interaction based recurrent convolutional neural network to learn the fine-grained correlations between the words in users' tweets and the health concepts. And 3) the recipe recommendation system aims to jointly consider the retrieved recipe candidates and users' health features to accomplish the personalized food recommendation. Towards this end, we present a category-aware hierarchical memory network to learn the close correlations among users with the same health tags, the relations among the recipes with similar nutritive values, and the interactions between users and recipes. Specifically, the recommender divides the users and the recipes into different categories according to their health tags and nutritive values, respectively. And then it leverages the category-level and recipe-level matching scores to learn the inter-category difference and the intra-category similarity for better food recommendation. Finally, the healthy recipes for users are output by the recommender. To justify our model, we construct 
two large-scale food-related datasets: user health profiling dataset from Weibo, and a health-aware food recommendation dataset. Extensive experiments demonstrate the superiority of our models in two tasks: user health profiling and health-aware food recommendation. Besides, we also develop a demo \cite{Jiang2019Market2Dish} to testify the effectiveness and efficiency of the whole scheme.

To sum up, the main contributions of our work are threefold:
\begin{itemize}
\item To the best of our knowledge, this is the first work on personalized health-aware food recommendation, mapping the ingredients displayed in the market to the dishes eaten at home. In addition, we release two high-quality food-related datasets and the involved codes\footnote{\url{https://github.com/WenjieWWJ/FoodRec}.} to facilitate the research community in this field. 
\item We present a word-class interaction based text classification model to profile the users' health conditions via their sparse health-related information posted on the social networks.
\item We propose a novel category-aware hierarchical memory network based food recommendation system to learn the health-aware user-recipe interactions.
\end{itemize}

\section{Related Work}
 
%
%

Due to the significance of food to human life and health, extensive research efforts have been dedicated to the food-related study. According to the latest food survey \cite{Weiqing2018A}, the research in food computing falls into five main tasks, containing perception, recognition, retrieval, recommendation, and monitoring. In particular, food perception \cite{min2017delicious, Weiqing2018you, ofli2017saki} studies how people perceive food from its characteristics while food recognition \cite{kagaya2014food,wu2016learning,yang2010food,Pouladzadeh:2017:MMR:3119899.3063592,min2019ingredient,an2017pic2dish} aims to recognize and detect the categories or ingredients of the meals. Differently, some work also focuses on the recipe modification based on the original recipes \cite{shidochi2009finding, blansche2010taaable} Besides, food retrieval \cite{salvador2017learning} comprises visual image-based retrieval \cite{kitamura2008food}, textual recipe-based retrieval \cite{wang2008substructure} and cross-modal recipe-image retrieval \cite{chen2018deep}. And food recommendation \cite{trattner2017investigating,ge2015health} leverages multi-faceted information to recommend healthy and delicious food to users. By contrast, food monitoring \cite{farseev2017tweet,sanjo2017recipe} is intended to analyze various health-related information, monitor and predict the public health based on the massive data from the social media. For this work, food recommendation, retrieval and monitoring are three highly relevant research directions.

%
%
\subsection{Food Recommendation}

With the explosive growth of data on the Internet, recommendation systems \cite{Wang2019Neural, wang2020denoising, Li2019Routing, cheng2014just, Cheng2016on} have been proven effective to alleviate the overloaded information. According to the recent survey \cite{trattner2017food}, the studies in the food recommendation can be divided into five categories, namely content-based, collaborative filtering-based, context-aware, hybrid-based, and health-aware food recommendation. 
To be more specific, content-based approaches \cite{freyne2010intelligent} focus on the recipes, including the ingredients and food images; whereas collaborative filtering-based ones \cite{ge2015using} leverage the classic collaborative filtering algorithms \cite{herlocker1999algorithmic} to achieve food recommendation, such as Singular Value Decomposition \cite{harvey2013you} and Matrix Factorization (MF) \cite{ge2015using}. For context-aware methods, they \cite{kusmierczyk2015temporality} explored the value of rich context about users in food recommendation, such as gender, hobbies, and culture. Moreover, hybrid-based approaches \cite{ge2015using} usually integrate several existing methods for the recipe recommendation. However, all these strategies focus on recommending recipes based on the historical records of users to satisfy their preferences. Recently, incorporating healthiness into the food recommendation has attracted extensive research attention in this community \cite{Jiang2019Market2Dish}. For example, Ge \textit{et al.} \cite{ge2015health} integrated the calorie consumption into the food recommendation and Markus \textit{et al.} \cite{rokicki2018impact} estimated the influence of various recipe features on health-aware recipe recommendation. However, existing health-aware approaches cannot profile the user health and make personalized food recommendation at the ingredient level and recipe level due to the lack of large-scale datasets.

%
%
\subsection{Food Retrieval}

Food-related data are usually presented in multiple modalities, including textual recipes, visual food images, and cooking videos. According to the retrieval types, the existing food retrieval methods comprise three categories: textual recipe-based retrieval \cite{wang2008substructure}, visual image-based retrieval \cite{kitamura2008food}, and cross-modal recipe-image retrieval \cite{salvador2017learning, harashima2017cookpad, chen2017cross, carvalho2018cross}. The textual recipe-based retrieval is designed to retrieve recipes by the textual recipe query whereas the visual image-based retrieval focuses on the understanding of the image query. Lately, many efforts have been dedicated to retrieving the textual recipe based on an image query, namely cross-modal recipe-image retrieval. For instance, Chen \textit{et al.} \cite{chen2017cross_mm} leveraged rich food attributes to retrieve the textual recipe given the food image as a query and achieved promising results. Different from the aforementioned methods, the recipe retrieval in this work is designed to first recognize the ingredients from the micro-videos taken in the market, and then retrieve recipes from the recipe dataset based on the captured ingredients.

%
%
%
%
\subsection{Food Monitoring}

According to the recent global digital report issued by \textit{We Are Social}\footnote{\url{https://wearesocial.com/}.}, there are over 4.39 billion Internet users, including 3.48 billion social media users in the world today. And it is an inexorable trend that the Internet users are spending more time on social networks. Therefore, large-scale personal data are accumulated on the Internet, providing rich sources for the prediction and analysis of public health. Indeed, many efforts \cite{farseev2017tweet, abbar2015you} have been dedicated to investigating the public health issues, such as the national obesity and diabetes \cite{abbar2015you}. However, the existing work is limited in that it only studies some food-related patterns based on the data from graph-based social networks \cite{feng2019graph, Feng2018learning}, such as the food consumption patterns \cite{mejova2015foodporn} or the diabetes rates in different regions \cite{de2016characterizing}. In this work, we turned to profile the user health based on the shared personal information in the social media and thereafter generated the personalized health-aware food recommendation.

Indeed, the task of user health profiling has been converted into a multi-label text classification problem by defining many user health tags. Therefore, this work is also closely related to many classic text classification methods, such as Fast Text Classifier (FastText) \cite{joulin2016bag}, Convolutional Neural Networks (TextCNN) \cite{kim2014convolutional}, Recurrent Convolutional Neural Network (RCNN) \cite{lai2015recurrent}, and Hierarchical Attention Networks (HAN) \cite{yang2016hierarchical}. Although great success has been achieved in text classification, especially with the rise of the data-driven deep neural models, these representative methods are impotent in this task. Because they usually learn a latent vector representation of the text content, and then calculate the probability of each class by projecting the latent vector presentation with a fully-connected (FC) layer. However, the health-related user information on the social networks is extremely sparse, thus learning high-level user representation from many tweets may be easily overwhelmed by the noise.

%
%

\begin{figure*}
\includegraphics[scale=0.4]{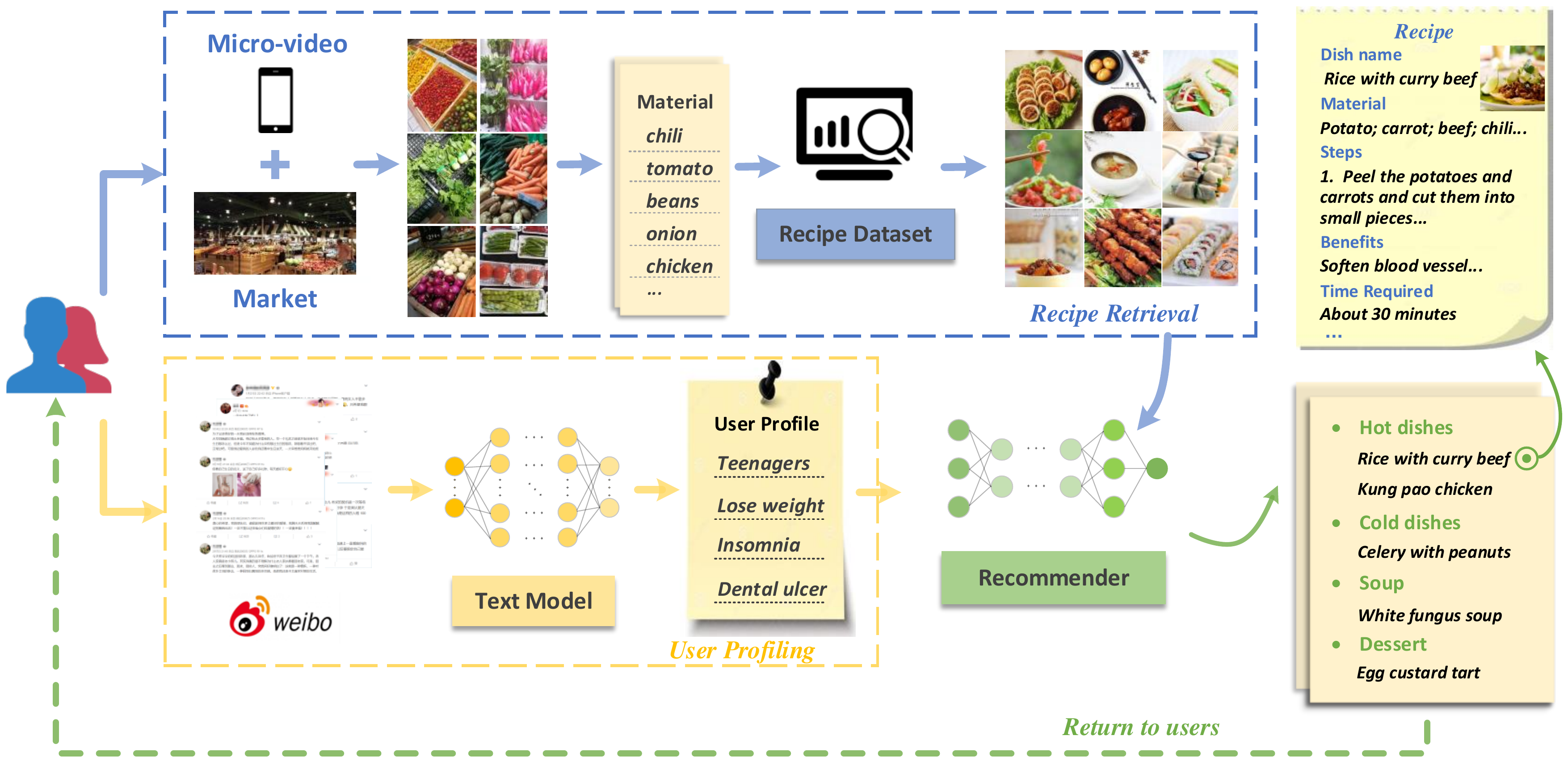}
\caption{Schematic illustration of our proposed model, comprising recipe retrieval, user health profiling, and health-aware recommender model. Ultimately, the personalized recipes with the cooking instructions are returned to users.}
\label{Figure2}
\end{figure*}

%
%
%
%
%
%
\section{Method}

In this paper, our proposed health-aware food recommender system aims to profile the user health, and thereafter recommend the personalized healthy food based on the ingredients they could buy from the market. As illustrated in Figure \ref{Figure2}, the whole scheme is divided into three components according to the roles, namely recipe retrieval, user health profiling, and personalized health-aware recipe recommendation. In this section, we will detail them one by one. 

%
%
%
%
\subsection{Recipe Retrieval}

The objective of this component is to acquire ingredients available to users, and then retrieve recipe candidates from large-scale recipe datasets. In the proposed framework, there are many ways to input ingredients for users, including selecting from the candidates, voice input, and micro-video input. In this work, we explore the possibility of micro-video input due to its higher difficulty. To this end, we sample images (frames) from the micro-videos taken by users in the market, recognize ingredients in the images via the multi-label image classification, and ultimately retrieve available recipes from a large-scale recipe dataset. In particular, we construct a large-scale dataset by human annotation, in which many ingredients are recorded by numerous images captured from the end markets. Based on our analyses and statistics on the ingredient images, there are almost 80 kinds of common ingredients, including various vegetables, seafood, and meat. Notably, an image usually contains multiple ingredients, and thus the ingredient recognition is essentially a task of multi-label image classification. Considering that lots of deep neural networks have achieved promising performance in image classification, we incorporate a pre-trained Inception-v3 Net \cite{szegedy2016rethinking} to tackle this multi-label image classification problem due to its high accuracy. Upon experiments, Inception-v3 Net achieves promising accuracy up to 95.5\%, which extremely facilitates the retrieval of available recipes from the recipe dataset. In the light of this, we obtain extensive recipe candidates based on the available ingredients. From the experiments, we find that micro-video input is also a feasible way to input ingredients for users.

%
%
%
%
\subsection{User Health Profiling}

As introduced in \cite{Weiqing2018A}, the online social networks with billions of users have provided extensive user data for the food-related prediction and monitoring. And as for user profiling, some people have proposed special ways to learn user behavior profiles in social networks \cite{zhao2016user, iglesias2011creating, yi2019privacy}. In this work, we design a supervised neural model to profile the user health by the rich user generated content distributed over the social networks. The experts in the health domain jointly consider various factors, such as gender, age, and common diseases, and then help us propose many health tags to characterize the common identities and health conditions of different people, like teenager, fitness enthusiast, and the diabetic. Notably, each user has at least one health tag (related to the job or age) and most of them have multiple ones. Therefore, the task of user health profiling will be converted into a multi-label classification problem: given the user generated content on the social networks, our proposed model will classify the user into one or several classes (\textit{i.e.}, different user health tags).

%
%

\begin{figure*}
\includegraphics[scale=0.75]{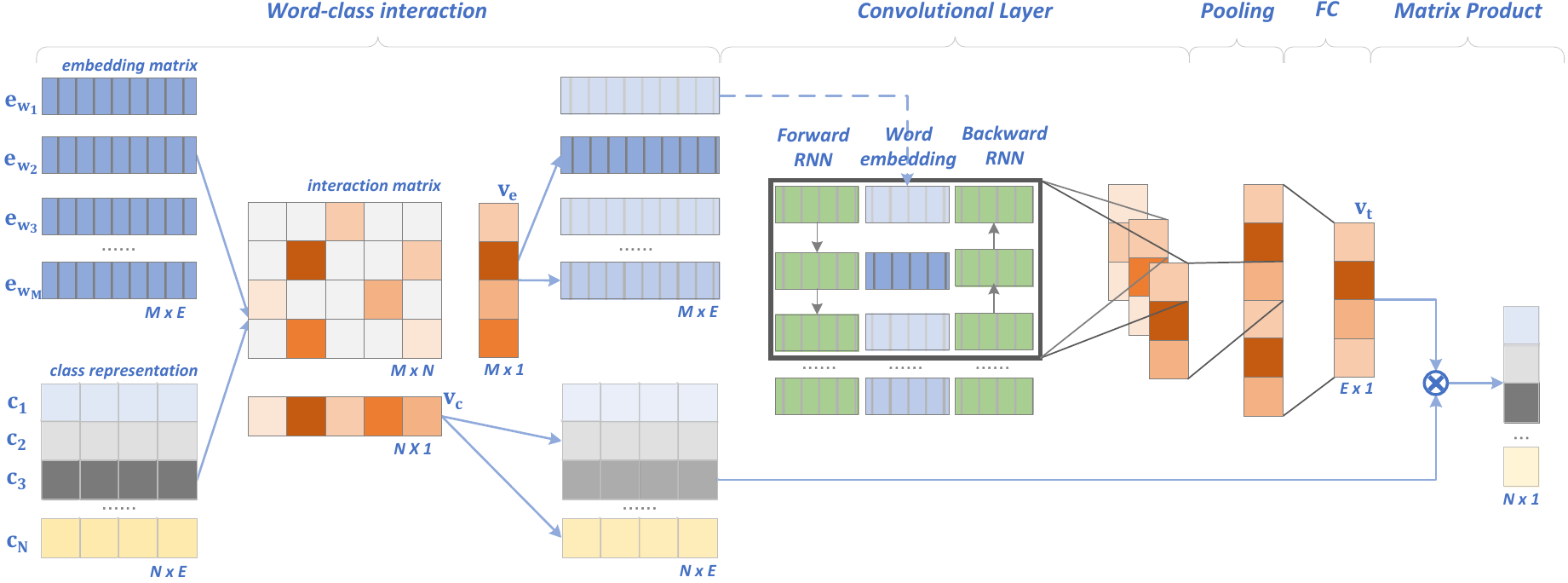}
\caption{Schematic illustration of our proposed WIRCNN. WIRCNN incorporates an interaction mechanism to learn the fine-grained correlations between the words and the classes, and leverages Bi-RNN and CNN to distill the textual contents for more distinguishable features.}
\label{Figure3}
\end{figure*}

%
%

To learn the correlation between the sparse health-related information and the health tags, we propose a Word-class Interaction based Recurrent Convolutional Neural Network, named WIRCNN for short, as shown in Figure \ref{Figure3}. Motivated by the phenomenon that only several keywords in the tweets have direct correlation with the health tags, we introduce the word-class interaction \cite{wang2015learning} to learn the fine-grained matching relations between the keywords and the classes. Thenceforth, WIRCNN leverages Bidirectional Recurrent Neural Networks (Bi-RNN) and Convolutional Neural Network (CNN) to encode the weighted word embeddings, and then outputs the probabilities over the classes via the matrix multiplication.

Formally, given the personal information and many tweets of a user, WIRCNN first concatenates them into a sequence of ordered tokens $\{w_1, w_2, ..., w_M\}$ by several special separators (\textit{e.g.}, using \_eos\_ to represent the end of sentence). Note that personal information such as age is putted before the tweets. We regard the matrix $\mathbf{W_c} \in \mathbb{R}^{N \times E}$ as the class representation, where $N$ and $E$ refer to the number of classes and the feature size, respectively. It is worth noting that the aforementioned FC layer in the existing methods can be interpreted as the class representation $\mathbf{W_c}$, and it essentially estimates the similarity between the latent vector representations of textual contents and the class representations via the matrix multiplication \cite{press2016using}. Thereafter, WIRCNN embeds the given tokens into an embedding matrix $\mathbf{W_e} \in \mathbb{R}^{M \times E}$, and computes the interaction matrix $\mathbf{I} \in \mathbb{R}^{M \times N}$ by the following equations,
%
%
\begin{equation}
\left \{
\begin{aligned}
&\mathbf{W_e}=[\mathbf{e_{w_1}}, \mathbf{e_{w_2}}, ..., \mathbf{e_{w_M}}]^T,\\
&\mathbf{I} = \mathbf{W_e}\mathbf{W_c^T},
\end{aligned}
\right .
\end{equation}
%
%
where $M$ is the number of given tokens and $\mathbf{e_{w_i}}$ denotes the embedding of the token $w_i$. Then the max pooling function is respectively applied to two different dimensions of $\mathbf{I}$. In the light of this, we can get two vectors $\mathbf{v_e} \in \mathbb{R}^{M}$ and $\mathbf{v_c} \in \mathbb{R}^{N}$. Based on these two vectors, the matrices $\mathbf{W_c}$ and $\mathbf{W_e}$ are weighted as follows,
%
%
\begin{equation}
\left \{
\begin{aligned}
&\mathbf{W^{'}_e} = \mathbf{v_e} \odot \mathbf{W_e},\\
&\mathbf{W^{'}_c} = \mathbf{v_c} \odot \mathbf{W_c},
\end{aligned}
\right .
\end{equation}
%
%
where $\odot$ refers to the element-wise multiply, $\mathbf{W^{'}_e}$ and $\mathbf{W^{'}_c}$ denote the weighted matrices of $\mathbf{W_c}$ and $\mathbf{W_e}$, respectively. 

Afterwards, WIRCNN encodes the weighted word embeddings $\mathbf{W^{'}_e}$ via a Bi-RNN step by step, and concatenates the bidirectional hidden states with the weighted word embedding at each step for the convolutional layer. In the convolutional layer, WIRCNN uses multiple filters with varying window sizes to obtain the text features, and then applies a max-overtime pooling operation \cite{collobert2011natural} over each feature map, which takes the highest feature value as the representation of this feature map. At last, a FC layer is used to project features from the pooling layer into a high-dimensional space in the feature size $E$. And the probability $\mathbf{p}$ over all classes is calculated as,
%
%
\begin{equation}
\begin{aligned}
&\mathbf{p} = Sigmoid(\mathbf{v_t^T}\mathbf{W^{'}_c}),
\end{aligned}
\end{equation}
%
%
where $\mathbf{v_t}$ is the text feature from the FC layer, $\mathbf{W^{'}_c}$ refers to the weighted class representation, and the Sigmoid function maps the feature values into the interval $[0, 1]$. Ultimately, a cross-entropy loss is applied to optimizing the whole neural model.

%
%
\begin{figure*}
\centering
\includegraphics[scale=0.8]{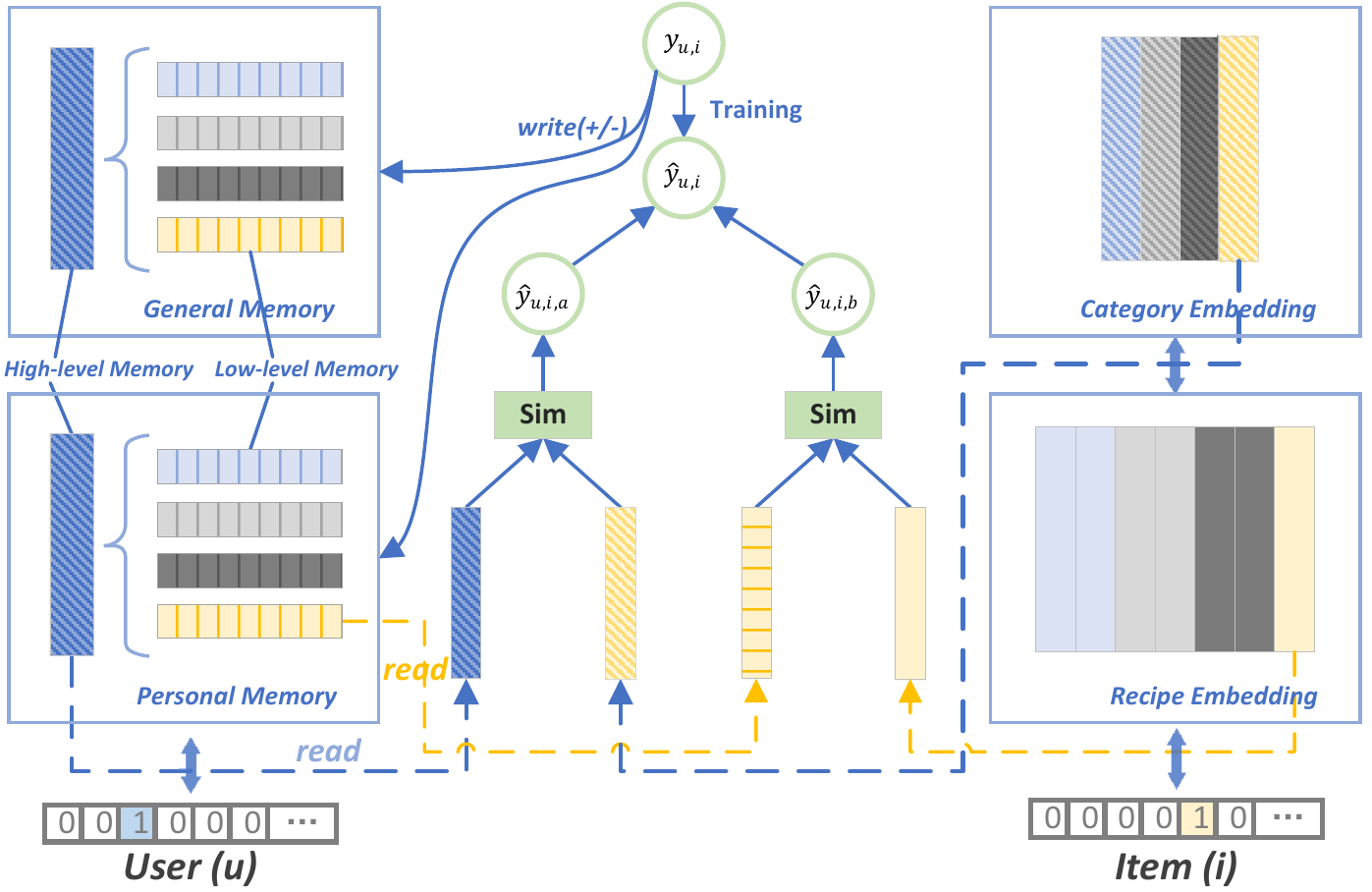}
\caption{The proposed recommender, consisting of four components: general memory, personal memory, category embedding, and recipe embedding. Given a user $u$ and a recipe $i$, the recommender first obtains the personal and general memories of the user $u$, and the recipe and category embeddings of the recipe $i$. Thereafter, it calculates the matching score hierarchically.}
\label{Figure4}
\end{figure*}
%
%
%
%
%
%
\subsection{Health-aware Recipe Recommendation}

The recommendation task in this component can be formulated as: given a user $u$ and the available recipe candidates $\{r_1, r_2, ..., r_{N_r}\}$, the recommender ranks these recipes based on the health tags of the user $u$ and the nutritive value of the recipes. For each user, we construct the positive samples $\mathcal{Y^+}=\{r_1^p, r_2^p, ..., r^p_{N_p}\}$ and the negative ones $\mathcal{Y^-}=\{r_1^n, r_2^n, ..., r^n_{N_n}\}$ from the recipe candidates according to the food-related health tips. Thereafter, the recommender can be trained by the health-aware user-recipe interactions.

In fact, close correlations not only exist among the users with similar health tags, but also the recipes sharing the same nutritive values. Intuitively, the users under the similar health conditions usually have similar diet habits, in other words, dishes made from the similar ingredients will be suitable for the same class of users. To leverage these correlations to improve the performance of food recommendation explicitly, we propose a category-aware hierarchical memory network to learn the intra-category similarity and the inter-category difference. All users fall into $N$ categories with different health tags, and the recipes are also divided into $N_c$ categories for different health needs, such as low calorie and nutritional supplements. Notably, each user and recipe could belong to multiple categories due to the multiple health tags or various nutritive values.

As illustrated in Figure \ref{Figure4}, the proposed recommender includes four components, namely general memory, personal memory, category embedding, and recipe embedding. In particular, recipe embedding and category embedding are utilized to encode the recipes and $N_c$ recipe categories into vectors, respectively. In addition, each user has a personal memory while a health tag corresponds to a general memory. The personal memory and the general one share the same internal structure, including a high-level memory vector and $N_c$ low-level memory vectors. Intuitively, the high-level memory vector records the health-aware preference of users for the recipe categories while the low-level ones remember the preference for each recipe in $N_c$ categories. In particular, the high-level memory corresponds to the category embedding and the low-level one corresponds to the recipe embedding. The general memory is applied to learning the common characteristics among users with the similar health conditions. For instance, the users who are losing weight should eat some low-fat food.

Formally, given a user $u$ and a recipe $i$, the recommender first obtains the personal memory of the user $i$, the recipe embedding $\mathbf{v^r_i}$, and the category embedding $\mathbf{v^c_i}$ of the recipe $i$. Afterwards, we calculate the score $\hat{y}_{u,i}$ for the user $u$ and the recipe $r$ as,
%
%
\begin{equation}
\begin{aligned}
\hat{y}_{u,i} = \alpha \times Sim(\mathbf{v^h_u}, \mathbf{v^c_i}) + (1 - \alpha) \times Sim(\mathbf{v^{l,i}_u}, \mathbf{v^r_i})),\\
\end{aligned}
\end{equation}
%
%
where $\mathbf{v^h_u}$ refers to the high-level memory vector of the user $u$, $\mathbf{v^{l,i}_u}$ denotes the low-level memory vector of the user $u$ regarding the category of the recipe $i$, and $\alpha$ is a hyper parameter to adjust the contribution of the high-level and low-level similarities. As to the $Sim(\mathbf{a}, \mathbf{b})$ function, we have tried several operations to calculate the similarity between $\mathbf{a}$ and $\mathbf{b}$, such as cosine similarity, dot product, and multi-layer perceptron (MLP). Besides, it is worth noting that if the recipe $i$ belongs to multiple categories, the category embedding $\mathbf{v^c_i}$ and the low-level personal memory vector $\mathbf{v^{l,i}_u}$ will be computed by taking the mean value of the corresponding vectors.
Finally, a binary cross entropy loss is applied to optimizing the recommender, which can be formulated as,
%
%
\begin{equation}
\begin{aligned}
\ell = -\sum_{u\in \mathcal{U}} \sum_{i\in \mathcal{Y^+}} y_{u,i} \log{\hat{y}_{u,i}} - \sum_{u\in \mathcal{U}} \sum_{i\in \mathcal{Y^-}} (1-y_{u,i})\log(1-\hat{y}_{u,i}),\\
\end{aligned}
\end{equation}
%
%
where $\mathcal{U}$ represents the set of all users, $y_{u,i}$ is the label of the recipe $i$ for the user $u$. Meanwhile, $y_{u,i}$ is 1 if the recipe $i$ is a positive sample, otherwise, 0. 

In addition, we define a write operation to update the personal memory and the general memory dynamically, and then leverage the general memory to update the corresponding personal memories with the same health tags at a certain frequency in the training. This operation will change the preference of the user $u$ for the recipe $i$ explicitly and leverage some common characteristics in the general memory to modify the preference of the user $u$. Specifically, assuming that the user $u$ has a health tag $t$ corresponding to the general memory $g_t$, the general memory $g_t$ will be updated through the following equation, 
%
%
\begin{equation}
\left \{
\begin{aligned}
&\mathbf{v^h_t} \leftarrow \mathbf{v^h_t} + z_{u,i}\beta^h\mathbf{v^c_i}, \\
&\mathbf{v^{l,c}_t} \leftarrow \mathbf{v^{l,c}_t} + z_{u,i}\beta^l\mathbf{v^r_i},\\
\end{aligned}
\right .
\end{equation}
%
%
where $\mathbf{v^h_t}$ denotes the high-level memory vector of $g_t$, $\mathbf{v^{l,c}_t}$ refers to the low-level one of $g_t$ corresponding to the category $c$ of the recipe $i$, $\beta^h$ and $\beta^l$ are hyper parameters, $\mathbf{v^c_i}$ and $\mathbf{v^r_i}$ are the category embedding and the recipe embedding of the recipe $i$, respectively. In addition, $z_{u,i}$ equals to 1 if $i \in \mathcal{Y^+}$, otherwise, -1. Moreover, the personal memory of the user $u$ is also updated by $\mathbf{v^c_i}$ and $\mathbf{v^r_i}$, similar to the general memory $g_t$. Intuitively, the write operation will increase/decrease the similarity between the memories of the user $u$ and the embeddings of the recipe $i$, which changes the preference of the user $u$ for the recipe $i$ explicitly. Notably, if a user has multiple health tags, all related general memories will be updated in this manner. Subsequently, the personal memory of the user $u$ can be updated by the general memory $g_t$:
%
%
\begin{equation}
\left \{
\begin{aligned}
&\mathbf{v^h_u} \leftarrow \mathbf{v^h_u} + \lambda^h\mathbf{v^h_t}, \\ 
&\mathbf{v^{l}_u} \leftarrow \mathbf{v^{l}_u} + \lambda^l\mathbf{v^l_t},\\
\end{aligned}
\right .
\end{equation}
%
%
where $\lambda^h$ and $\lambda^l$ are hyper parameters, $\mathbf{v^h_u}$ and $\mathbf{v^l_u}$ denote the high-level and low-level personal memory vectors of the user $u$, respectively. All low-level personal memories of the user $u$ are updated by the corresponding low-level general memories. In particular, if a user with several health tags has more than one general memory, the personal memory will be updated by the average of these general memories.

%
%
%
%

%
%
%
\section{Experiments}
%
%

%
%
%
%
\subsection{Data Collection}

To achieve the personalized health-aware food recommendation based on the available ingredients for users in the market, we constructed two food-related datasets: user health profiling dataset, and food recommendation dataset. And all the experimental data have been released to facilitate the research in the food recommendation domain. More details about the datasets and the specific cases can be found in our website\footnote{\url{https://healthawarerec.wixsite.com/foodrec}.}.


\subsubsection{User Health Profiling Dataset}

Nowadays, social networks have become an essential part in our daily life. Extensive users share their activities, feelings, and hobbies on these social network platforms every day, and these user-related information helps us to analyze users' identities, emotions, and even their health conditions. To profile the user health, we crawled extensive users' tweets along with their personal information (\textit{e.g.}, sex, age, and occupation)  from Weibo\footnote{\url{https://www.weibo.com}.}, one of the biggest social media platforms in China. Thereafter, many work for data cleaning and preprocessing was done to improve the quality of the dataset, such as the removal of duplicate and noisy tokens. In addition, we collected the health tags under the guidance of domain experts to cover the common health conditions of the Weibo users from multiple perspectives, such as age, and diseases. In particular, the health tags derived from the user's personal information mainly include age-related ones (i.e., school-age child, teen, middle-age users, elderly). And other tags obtained based on tweets are like losing weight, insomnia, and so on. Eventually, we acquired 64,657 Weibo users with 96 health tags (\textit{e.g.}, pregnant women, insomnia, hypertension, and obesity), and then we annotated users with the health tags based on their tweets and personal information from Weibo via a semi-automatic annotation method, integrating the keyword-based filter rules and human inspection. First, we retrieved the health-related tweets from the extensive tweets for each user by some health-related keyword matching rules, and then annotators assigned each user with several appropriate health tags based on the retrieved tweets and the user's personal information. In particular, we defined many health-related keywords such as slimming, exercise and disease, and selected the tweets containing these keywords for the convenient annotation. Besides, each user has a health tag at least, and usually has multiple ones. On average, each user has more than 40 tweets crawled from Weibo and 3.89 health tags.

\subsubsection{Food Recommendation Dataset}

Existing recipe websites provide us abundant food-related data for the research in the food domain. We crawled a large-scale Chinese recipe dataset from Meishijie\footnote{\url{http://meishij.net/}.}, consisting of over 6.6k recipes, of which each contains rich information, like the recipe name, ingredients, cooking instructions, dish pictures, benefits, cooking time, and cooking difficulty. In particular, Meishijie provides the benefits of each recipes, such as help digest, anti-diarrhea, and treatment of insomnia. To reduce the number of the category, we summarized all the recipes into 4 categories from the perspective of health (\textit{i.e.}, losing weight, health care, nutritional supplements, and disease recovery). It's worth noting that a recipe may belong to more than one category due to their multiple nutritional benefits. Besides, we acquired extensive healthy diet tips for people with different health conditions, for example the diabetes should eat more food rich in fiber and vitamin, such as carrot and celery, and meanwhile avoid sausage, lard oil and other ingredients rich in sugar and cholesterol. Based on the healthy diet tips, we could collect over 20 appropriate and inappropriate ingredients for the users with one kind of health tags. And then by matching the ingredients in recipes, we constructed lots of positive and negative health-aware dish samples from the crawled recipe dataset for each Weibo user. In detail, we collected 28,800 health-aware triples <health tag, suitable recipe, unsuitable recipe>. Each health tag corresponds 300 triples. 
Then, to simulate the ingredients that users could buy from the market, we assigned many ingredients for each Weibo user randomly to record the available ingredients for him/her.
Therefore, the positive and negative samples of the users with multiple health tags were filtered by the available ingredients, and then we acquired extensive user-recipe pairs from the perspective of health. Ultimately, the food recommendation dataset consists of 64,657 training samples, and each of them contains a Weibo user, the health tags, the available ingredients, about 340 positive recipes and 100 negative recipes for this user from the recipe dataset. 

%
%
%
%
\subsection{Experimental Settings}
%
%
\subsubsection{Hyper parameters}
In WIRCNN, the feature size $E$, the number of health tags $N$, and the length of users' Weibo tweets $M$ are 128, 96, and 390, respectively. Besides, the size of hidden states of the Bi-RNN was set as 128. As to the recommender, the numbers of users and recipes are 64,657, and 4,548, respectively. The category of recipes is 4. In addition, we chose inner product as the function $Sim(\mathbf{a}, \mathbf{b})$ due to the better experimental performance. The write operation is employed once for each user at each iteration. For the two neural models, we leveraged Adam as the optimizer with the learning rate initialized as 0.001. 

\subsubsection{Embedding Pre-training}
Lots of experiments have proven that parameter initialization plays a great role in many natural language processing and computer vision tasks \cite{mikolov2013distributed}. In addition, the health tags of users, the recipe images, and the recipe ingredients provide rich context information for the food recommendation. Therefore, we discussed the performance of two kinds of pre-training methods on food recommendation: item2vector \cite{barkan2016item2vec} and initialization by deep models. In this work, we used them to pre-train the latent embeddings of users and recipes. In particular, item2vector is derived from word2vector \cite{mikolov2013distributed}. The users and recipes were divided into many groups for the pre-training according to the predefined health tags and recipe categories. As to the deep model initialization, we designed a deep neural model to encode rich context information for the pre-training of the latent embeddings of users and recipes. The latent embeddings of users were calculated from their health tags by a FC layer while the latent embeddings of recipes were acquired by the concatenation of the visual features of recipe images and the textual features of recipe ingredients. In particular, a VGGNet-19 \cite{simonyan2014very} was applied to extracting the visual features of recipe images; meanwhile a TextCNN model was incorporated to distill the embeddings of ingredients and output the textual features of ingredients. Notably, the embedding of ingredients were randomly initialized in this model. Thereafter, we leveraged a MLP model to calculate the matching score between the latent embeddings of users and recipes. By training this deep neural model, we can obtain the latent embedding for each user and recipe. Ultimately, these pre-trained vectors were utilized to initialize the personal memory and the recipe embedding in our proposed recommender. It is worth noting that the high-level and low-level memory vectors of each user were initialized with the same latent embedding. Moreover, the general memory and the category embedding took the mean values of the corresponding personal memories and recipe embeddings, respectively.

%
%
\subsubsection{Baselines}
To evaluate our proposed models, we compared them with several state-of-the-art methods in two tasks, respectively.
Regarding the task of user health profiling, we compared the proposed WIRCNN with the following baselines: 
\begin{itemize}

\item \textbf{FastText} \cite{joulin2016bag} averages the word embeddings of textual content, and then predicts the probability of classification. 
\item \textbf{TextCNN} \cite{kim2014convolutional} leverages multiple filters to obtain textual features and outputs the probability distribution over all classes by a fully connected softmax layer. 
\item \textbf{RCNN} \cite{lai2015recurrent} employs a Bi-RNN to capture the contextual information and a max-pooling layer to acquire the key features. 
\item \textbf{HAN} \cite{yang2016hierarchical} incorporates the hierarchical attention network to learn the attention weights of textual features at the word level and the sentence level.
\item \textbf{EXplicit interAction Model (EXAM)} \cite{du2019exam} incorporates the interaction mechanism \cite{wang2015learning} into text classification and achieves promising performance.
\end{itemize}

As to the health-aware food recommendation, we incorporated several general recommendation frameworks as the baselines: 
\begin{itemize}

\item \textbf{MF} is the most popular collaborative filtering algorithm, conducting recommendation by calculating the inner product between the latent embeddings of the item and the user. 
\item \textbf{Generalized Matrix Factorization (GMF)} \cite{he2017neural} generalizes the inner product of MF towards a non-linear neural layer. 
\item \textbf{Neural collaborative filtering (NCF)} \cite{he2017neural} employs a MLP model to replace the inner product of MF with the aim of capturing more useful information between the latent embeddings of the user and the item. 
\item \textbf{Neural matrix factorization (NeuMF)} \cite{he2017neural} is a fusion model of GMF and NCF (MLP), concatenating the features from GMF and NCF (MLP) to output the probability by a non-linear neural layer. 
\end{itemize}
To learn more representative latent embeddings of users and items, we also computed the latent embeddings of all baselines by deep neural models with the same setting in Section 4.2.2. The difference is that the parameters of deep models in baselines for feature extraction can be optimized.

%
%
\subsubsection{Evaluation Metrics}

In the task of user health profiling, we employed ten-fold cross validation to estimate the performance of WIRCNN and the baselines. In addition, we adopted \textit{micro-precision} (dubbed as micro-P), \textit{micro-recall} (micro-R), \textit{micro-F1}, and \textit{macro-F1} to objectively evaluate their performance following the former work \cite{lai2015recurrent,kim2014convolutional}. In particular, micro-averaged scores (\textit{i.e.}, micro-P, micro-R and micro-F1) are calculated by the global confusion matrix, the sum of the confusion matrices of all categories; whereas macro-F1 is computed by first calculating the per-category F1 score and then taking the average of the F1 scores for all categories \cite{yang1999evaluation}. Micro-averaged scores give the equal weights to each sample-category pair from the micro perspective, and hence the samples belonging to more categories will occupy larger weights; whereas the macro-averaged ones treat each category equally at the macro level \cite{yang1999evaluation}. 

As to the food recommendation task, we employed \textit{leave-one-out} evaluation method, which is widely adopted in existing recommendation studies \cite{bayer2017generic, he2016fast, he2017neural}. In particular, for each user, we left a positive sample and 50 negative samples as the test data, and leveraged the remaining positive and negative samples for training. We ran all models 10 times and randomly chose different test data every time. To judge the proposed recommender, we utilized \textit{Hit Ratio} (HR), \textit{Normalized Discounted Cumulative Gain} (NDCG), and \textit{Area Under the Roc Curve} (AUC) to justify the performance \cite{wang2020click}. In the testing period, we selected the top-5 and top-10 recipes from the recommendation list for each user, and then respectively calculated the averaged HR@5, NDCG@5, HR@10, and NDCG@10 for all the testing users. Intuitively, AUC measures the probability that the recommender ranks a randomly chosen positive sample higher than a negative sample, and HR@k represents the hit ratio that a positive sample is ranked at the top-k positions while NDCG assigns higher weights to the hits at the top ranks. For all metrics, higher scores denote better performance.

%
%
\begin{table*}[]
\setlength{\abovecaptionskip}{0cm}
\setlength{\belowcaptionskip}{0cm}
\caption{Performance comparison between the baselines and WIRCNN regarding the task of user health profiling. We employ ten-fold cross validation and report mean$\pm$standard deviation. The p-values of the student t-test are much smaller than 0.05, indicating that WIRCNN is significantly superior.}
\label{table1}
\centering
\setlength{\tabcolsep}{2.8mm}{
\begin{tabular}{l|l|l|l|l}
\toprule
\textbf{Methods} & \textbf{Macro-F1} & \textbf{Micro-F1} & \textbf{Micro-P} & \textbf{Micro-R} \\ \hline
FastText & 0.1014$\pm$0.0023 & 0.6271$\pm$0.0036 & 0.8271$\pm$0.0076 & 0.5049$\pm$0.0059 \\ \hline
EXAM & 0.7322$\pm$0.0134 & 0.8795$\pm$0.0067 & 0.8914$\pm$0.0074 & 0.8680$\pm$0.0073 \\ \hline
RCNN & 0.8783$\pm$0.0143 & 0.9752$\pm$0.0009 & 0.9788$\pm$0.0027 & 0.9717$\pm$0.0023 \\ \hline
TextCNN & 0.8820$\pm$0.0120 & 0.9696$\pm$0.0012 & 0.9630$\pm$0.0025 & 0.9761$\pm$0.001 \\ \hline
HAN & 0.8372$\pm$0.0192 & 0.9612$\pm$0.0018 & 0.9751$\pm$0.0025 & 0.9473$\pm$0.0021 \\ \hline
\textbf{Ours} & \textbf{0.9017$\pm$0.0113} & \textbf{0.9784$\pm$0.0012} & \textbf{0.9811$\pm$0.0017} & \textbf{0.9762$\pm$0.0014} \\
\bottomrule
\end{tabular}}
\end{table*}
%
%

%
%
%
%
\subsection{Overall Performance}

\subsubsection{Evaluating the User Health Profiling}
Table \ref{table1} summarizes the performance of all baselines and WIRCNN with respect to several standard metrics on the task of user health profiling. Following prior work \cite{wang2018chat, Wang2019Neural, Feng2018learning}, we conducted significance test to evaluate the stability of the proposed method. From Table \ref{table1}, we can observe the following points: 1) WIRCNN consistently achieves the superior performance than all the baselines, especially under the metrics of Micro-F1 and Macro-F1. This reflects that interaction mechanism, Bi-RNN, and CNN in WIRCNN can capture more health-related user information from the Weibo tweets for the classification, demonstrating the effectiveness of our proposed WIRCNN on the sparse user health profiling dataset. 2) HAN outperforms the other baselines by attentively learning the textual representation at the word and sentence levels. Incorporating the success of the interaction mechanism in WIRCNN, we could conclude that attentively extracting textual features is really helpful in this task. And 3) FastText is the worst one among the baselines, probably because that it treats all the words equally in the user's tweets and hence incorporates much noise.

%
%
\begin{table*}[]
\setlength{\abovecaptionskip}{0cm}
\setlength{\belowcaptionskip}{0cm}
\caption{Performance comparison between the baselines and our proposed recommender on the task of health-aware food recommendation. We randomly change the test data, run all models 10 times, and then report mean $\pm$ standard deviation. The proposed recommender significantly outperforms the baselines according to the student t-test ($\text{p-value} < 0.05$).}
\label{table2}
\centering
\setlength{\tabcolsep}{0.9mm}{
\begin{tabular}{l|l|l|l|l|l}
\toprule
\textbf{Methods} & \textbf{HR@5} & \textbf{NDCG@5} & \textbf{HR@10} & \textbf{NDCG@10} & \textbf{AUC} \\ \hline
MF & 0.8393$\pm$0.0102 & 0.7491$\pm$0.0190 & 0.9095$\pm$0.0048 & 0.7717$\pm$0.0168 & 0.9195$\pm$0.0071 \\ \hline
GMF & 0.8331$\pm$0.0121 & 0.7380$\pm$0.0219 & 0.9059$\pm$0.0026 & 0.7618$\pm$0.0196 & 0.9301$\pm$0.0042 \\ \hline
NCF(MLP) & 0.7837$\pm$0.0113 & 0.6870$\pm$0.0137 & 0.8624$\pm$0.0093 & 0.7123$\pm$0.0111 & 0.9205$\pm$0.0047 \\ \hline
NeuMF & 0.8416$\pm$0.0086 & 0.7560$\pm$0.0076 & 0.9086$\pm$0.0051 & 0.7786$\pm$0.0060 & 0.9300$\pm$0.0047 \\ \hline
\textbf{Ours} & \textbf{0.9008$\pm$0.0012} & \textbf{0.8046$\pm$0.0017} & \textbf{0.9548$\pm$0.0014} & \textbf{0.8208$\pm$0.0033} & \textbf{0.9570$\pm$0.0056} \\ 
\bottomrule

\end{tabular}}
\end{table*}
%
%

%
%

\begin{figure}
  \centering 
  \hspace{-0.4in}
  \subfigure{
    \includegraphics[width=1.6in]{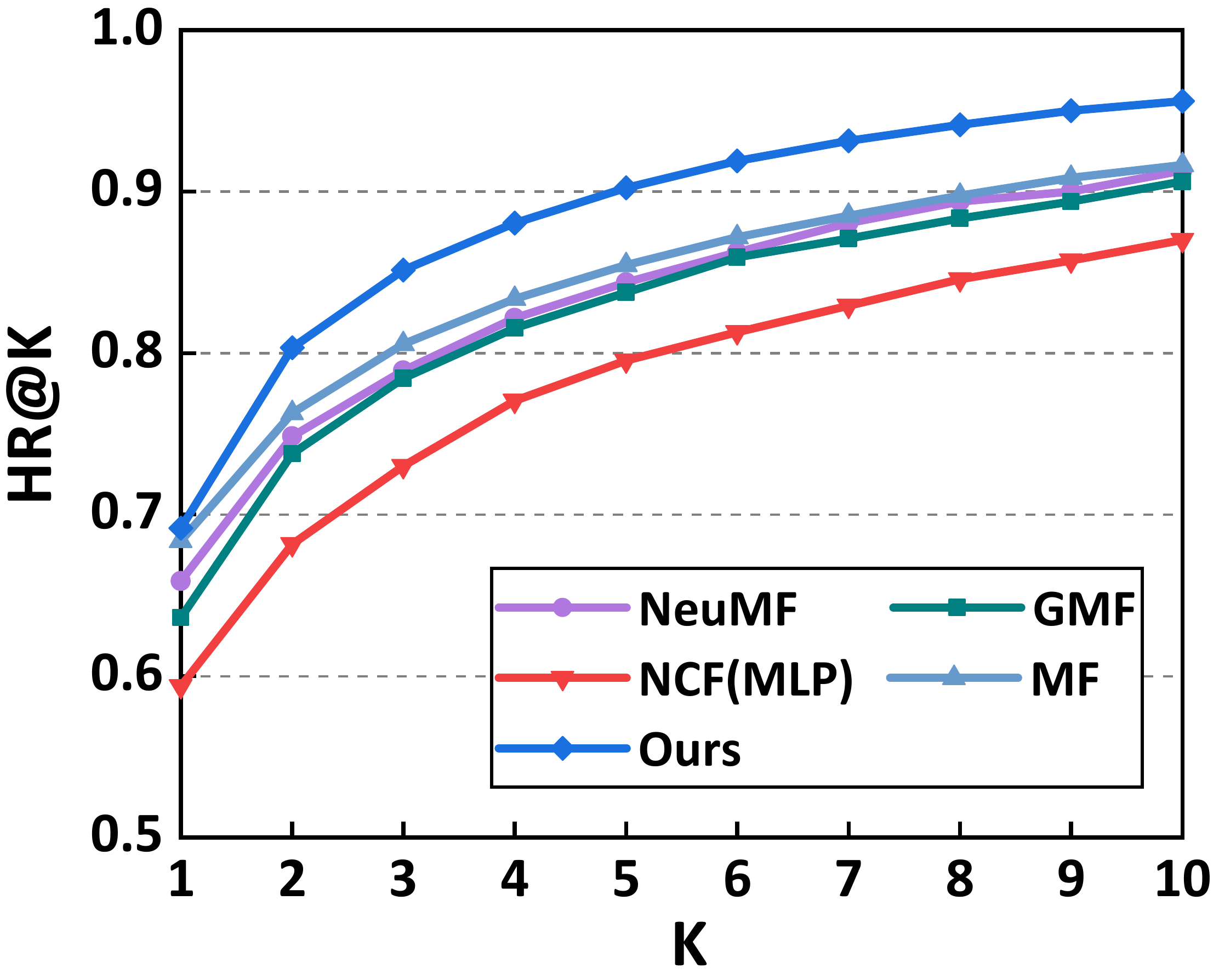}} 
  \subfigure{
    \includegraphics[width=1.6in]{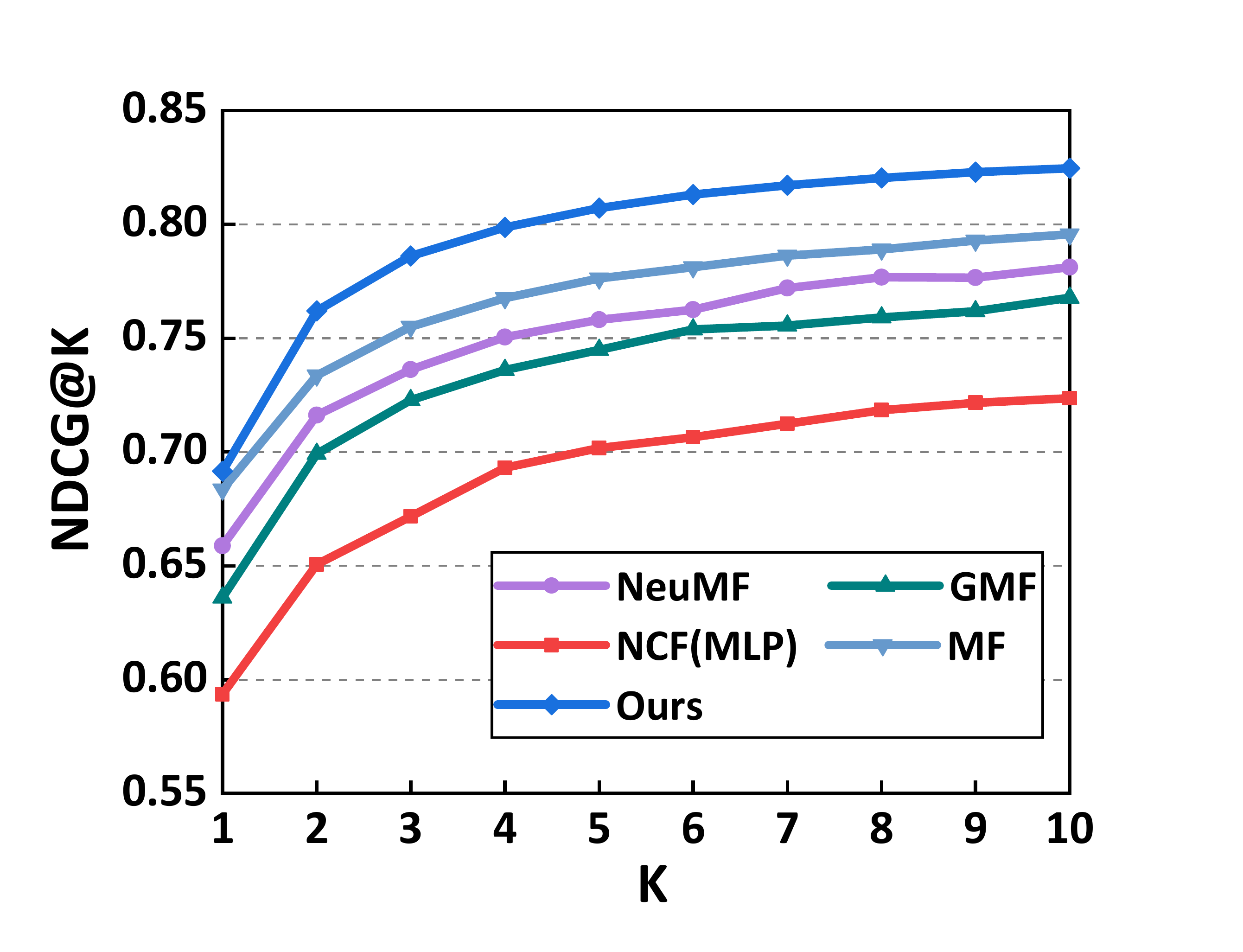}} 
  \vspace{-0.1in}
  \caption{Performance comparison with respect to Top-K item evaluation where K varies from 1 to 10.} 
  \label{Figure5}
\end{figure}
%
%
\subsubsection{Evaluating the Health-aware Food Recommendation}

The performance comparison between the baselines and our proposed recommender is presented in Table \ref{table2} and Figure \ref{Figure5}. Notably, the latent embeddings of users and recipes are obtained by the deep model in the baselines and our recommender. By contrast, we can have the following findings: 1) the proposed recommender significantly outperforms the baselines. Besides, in Figure \ref{Figure5}, the performance of our recommender keeps higher than the baselines regarding HR@k and NDCG@k when k varies from 1 to 10. This verifies that the recommender generates more appropriate recipes for the users by learning the similarity and difference at the category level. And 2) the fusion model NeuMF, combining GMF with NCF (MLP), yields better performance than the single model GMF and NCF (MLP). The observed results are consensus with the work in \cite{he2017neural}.

%
%
%
%

%
%
%
%
\subsection{Discussion}

%
%

\subsubsection{Utility of Pre-training}
To evaluate the utility of pre-training, we compared the performance of the proposed recommender and the best baseline MF with different pre-training methods. 
From the results in Table \ref{table3}, we can summarize the conclusions as follows: 1) The proposed recommender without the pre-training still surpasses the pre-trained MF, illustrating that the category-level correlations are relatively crucial in this task. Besides, the recommender leverages the fixed pre-training embeddings whereas MF employs the same deep model to extract the features of users and recipes dynamically. Therefore, the experimental results clearly show the superiority of our proposed recommender. 2) The features extracted by the proposed deep model reinforce MF and our recommender with better performance, which reflects that incorporating the rich context of recipes and users is significant. 3) MF is more sensitive to the user and recipe features extracted by the deep model than our proposed recommender. This demonstrates that the incorporation of the category-level information into our recommender partly reduces its dependence on the pre-training. And 4) the Item2vector initialization provides more category-level information about users and recipes for the proposed recommender and makes it achieve the best performance.

%
%
\begin{table}[]
\setlength{\abovecaptionskip}{0cm}
\setlength{\belowcaptionskip}{-0.0cm}
\caption{Performance of MF and the proposed recommender with different pre-training methods.}
\label{table3}
\setlength{\tabcolsep}{0.6mm}{
\begin{tabular}{l|l|l|l|l|l} 
\toprule
\multicolumn{2}{l|}{\textbf{Pre-training Methods}} & \textbf{HR@5} & \textbf{NDCG@5} & \textbf{HR@10} & \textbf{NDCG@10} \\ 
\hline
{MF} & No pre-training & 0.7195 & 0.6071 & 0.8470 & 0.6461 \\ 
\cline{2-6}
 & Item2vector & 0.8291 & 0.7382 & 0.8977 & 0.7606 \\ 
\cline{2-6}
 & \textbf{Deep model} & \textbf{0.8545} & \textbf{0.7763} & \textbf{0.9165} & \textbf{0.7956} \\ 
\hline
{Ours} & No pre-training & 0.8882 & 0.7984 & 0.9497 & 0.8184 \\ 
\cline{2-6}
 & \textbf{Item2vector} & \textbf{0.9111} & \textbf{0.8274} & \textbf{0.9606} & \textbf{0.8436} \\ 
\cline{2-6}
 & Deep model & 0.9025 & 0.8072 & 0.9561 & 0.8247 \\
\bottomrule
\end{tabular}}
\end{table}
%
%

%
%
\begin{table}[]
\vspace{0cm}
\setlength{\abovecaptionskip}{0cm}
\setlength{\belowcaptionskip}{-0.0cm}
\caption{The results of ablation test on WIRCNN and the recommender.}
\label{table4}
\centering
\setlength{\tabcolsep}{0.8mm}{
\begin{tabular}{l|l|l|l|l|l}
\toprule
\multicolumn{2}{l|}{\textbf{WIRCNN}} & \multicolumn{1}{c|}{\textbf{Macro-F1}} & \multicolumn{1}{c|}{\textbf{Micro-F1}} & \multicolumn{1}{c|}{\textbf{Micro-P}} & \textbf{Micro-R} \\ \hline
\multicolumn{2}{l|}{No interaction} & 0.843 & 0.977 & 0.981 & 0.973 \\ \hline
\multicolumn{2}{l|}{No Bi-RNN} & 0.649 & 0.941 & 0.957 & 0.927 \\ \hline
\multicolumn{2}{l|}{\textbf{Ours}} & \textbf{0.881} & \textbf{0.981} & \textbf{0.983} & \textbf{0.978} \\ \hline \hline
\multicolumn{2}{l|}{\textbf{\begin{tabular}[c]{@{}l@{}}Recommender\\ (Deep model)\end{tabular}}} & \textbf{HR@5} & \textbf{NDCG@5} & \textbf{HR@10} & \textbf{NDCG@10} \\ \hline
\multicolumn{2}{l|}{No general memory} & 0.8946 & 0.8052 & 0.9505 & 0.8235 \\ \hline
\multicolumn{2}{l|}{No category embedding} & 0.8227 & 0.7138 & 0.9080 & 0.7416 \\ \hline
\multicolumn{2}{l|}{\textbf{Ours}} & \textbf{0.9025} & \textbf{0.8072} & \textbf{0.9561} & \textbf{0.8247} \\ \hline \hline
\multicolumn{2}{l|}{\textbf{\begin{tabular}[c]{@{}l@{}}Recommender\\ (Item2vector)\end{tabular}}} & \textbf{HR@5} & \textbf{NDCG@5} & \textbf{HR@10} & \textbf{NDCG@10} \\ \hline
\multicolumn{2}{l|}{No general memory} & 0.8998 & 0.8231 & 0.9519 & 0.8401 \\ \hline
\multicolumn{2}{l|}{No category embedding} & 0.8767 & 0.7915 & 0.9332 & 0.8100 \\ \hline
\multicolumn{2}{l|}{\textbf{Ours}} & \textbf{0.9111} & \textbf{0.8274} & \textbf{0.9606} & \textbf{0.8436} \\ \bottomrule
\end{tabular}}
\end{table}

%
%
\subsubsection{Model Ablation}
To investigate the effectiveness of multiple components in our proposed WIRCNN and recommender, we conducted the ablation test on two models \cite{Nie2019multimodal}. Table \ref{table4} lists the results of the ablation test. Specifically, for WIRCNN, we removed the interaction mechanism and Bi-RNN, respectively. From Table \ref{table4}, we can observe that the performance drops significantly when removing the Bi-RNN or interaction mechanism, indicating that Bi-RNN is important for WIRCNN to capture the contextual information, and interaction mechanism significantly promotes WIRCNN regarding the F1 scores. As to the recommender, we evaluated the effectiveness of the general memory and the category embedding by removing them one by one. From Table \ref{table4}, we can find that both the general memory and the category embedding greatly improve the performance of the recommender with two kinds of per-training methods. 
In addition, the recommender without category embeddings performs similarly to the baselines. Therefore, we can conclude that the similarity and difference at the category level are the foundation of our superior performance.

%
%

\begin{figure*}
  \centering 
      \subfigure{
    \includegraphics[width=1.3in]{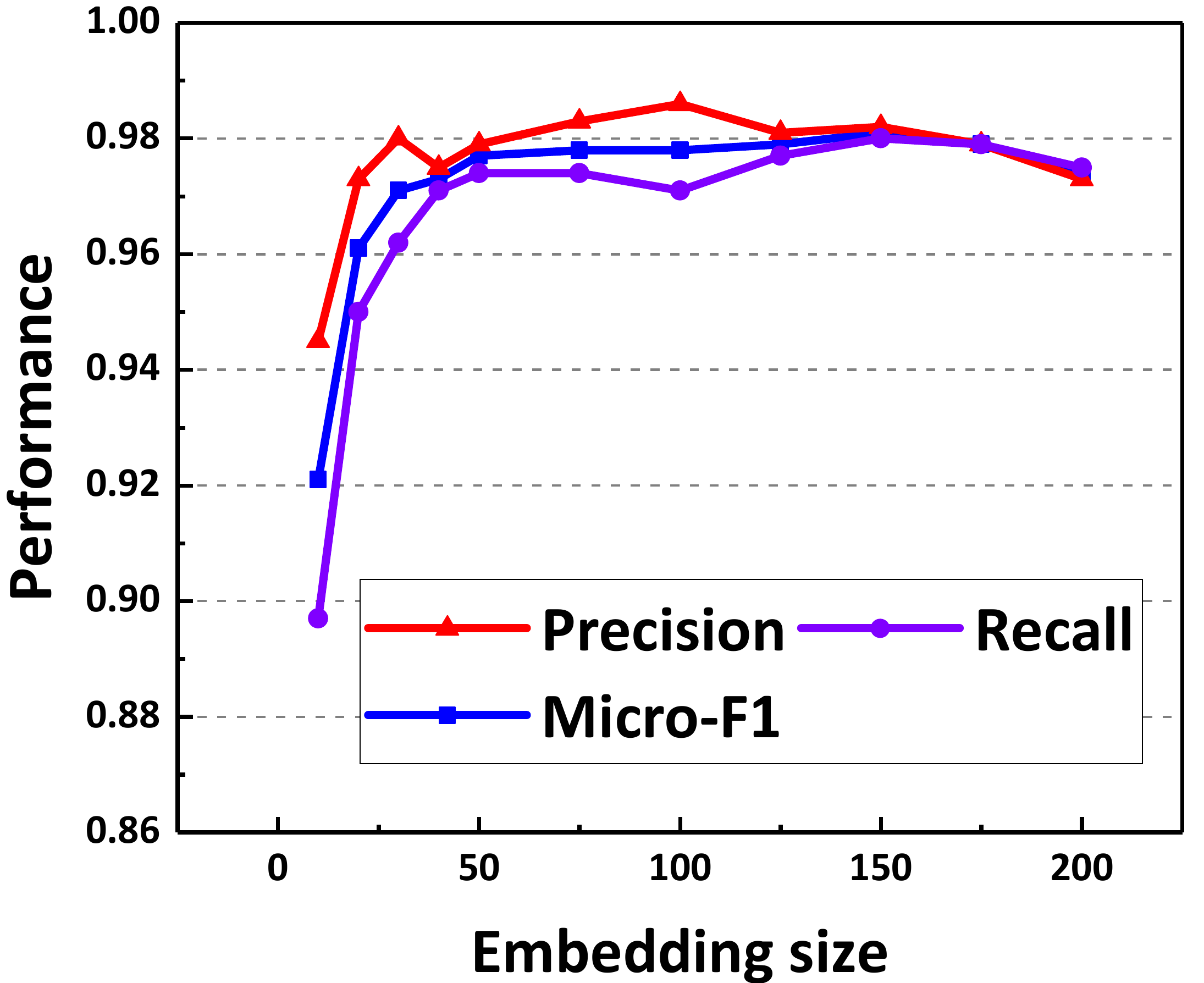}} 
  \subfigure{ 
    \includegraphics[width=1.3in]{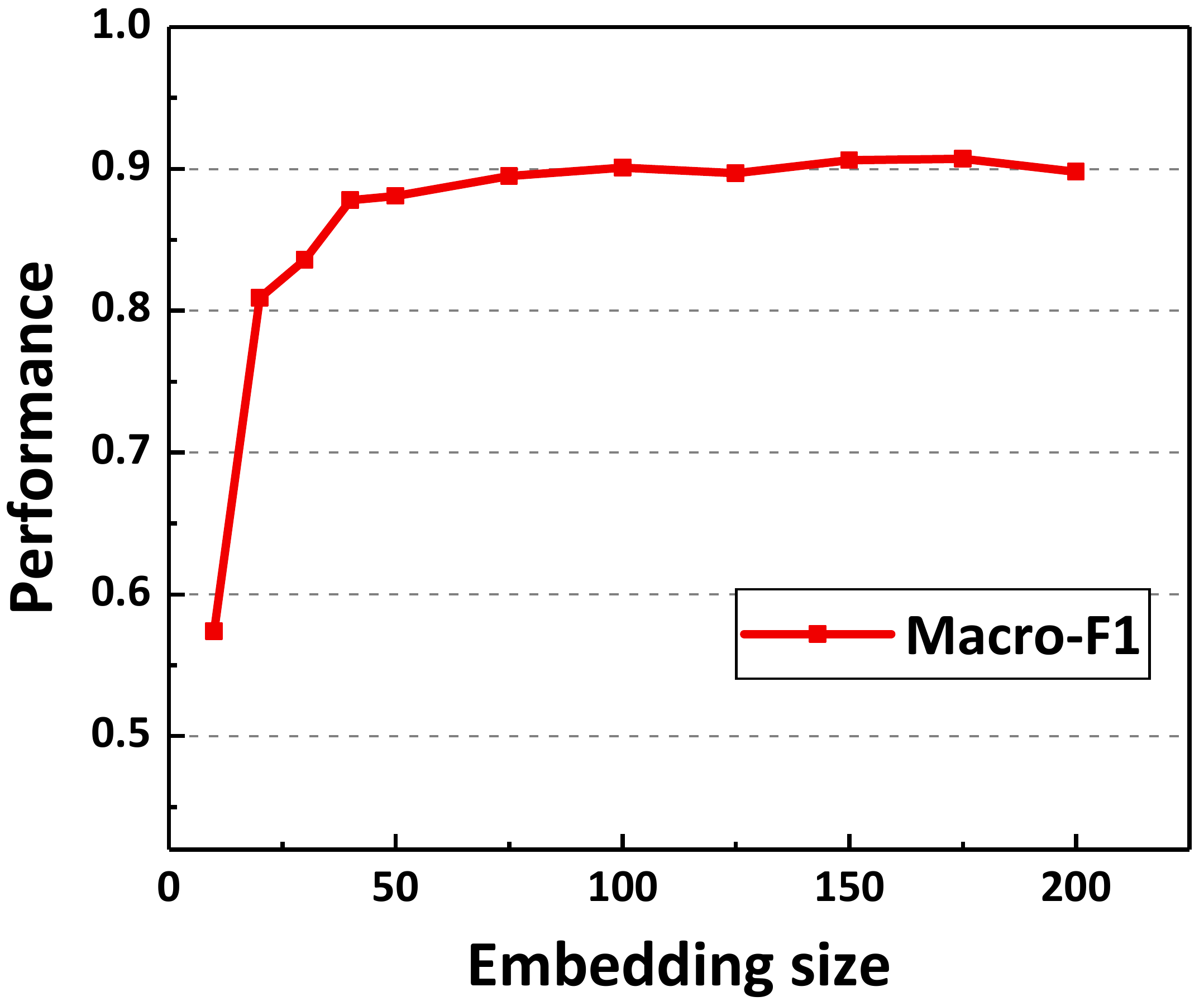}}
  \subfigure{
    \includegraphics[width=1.3in]{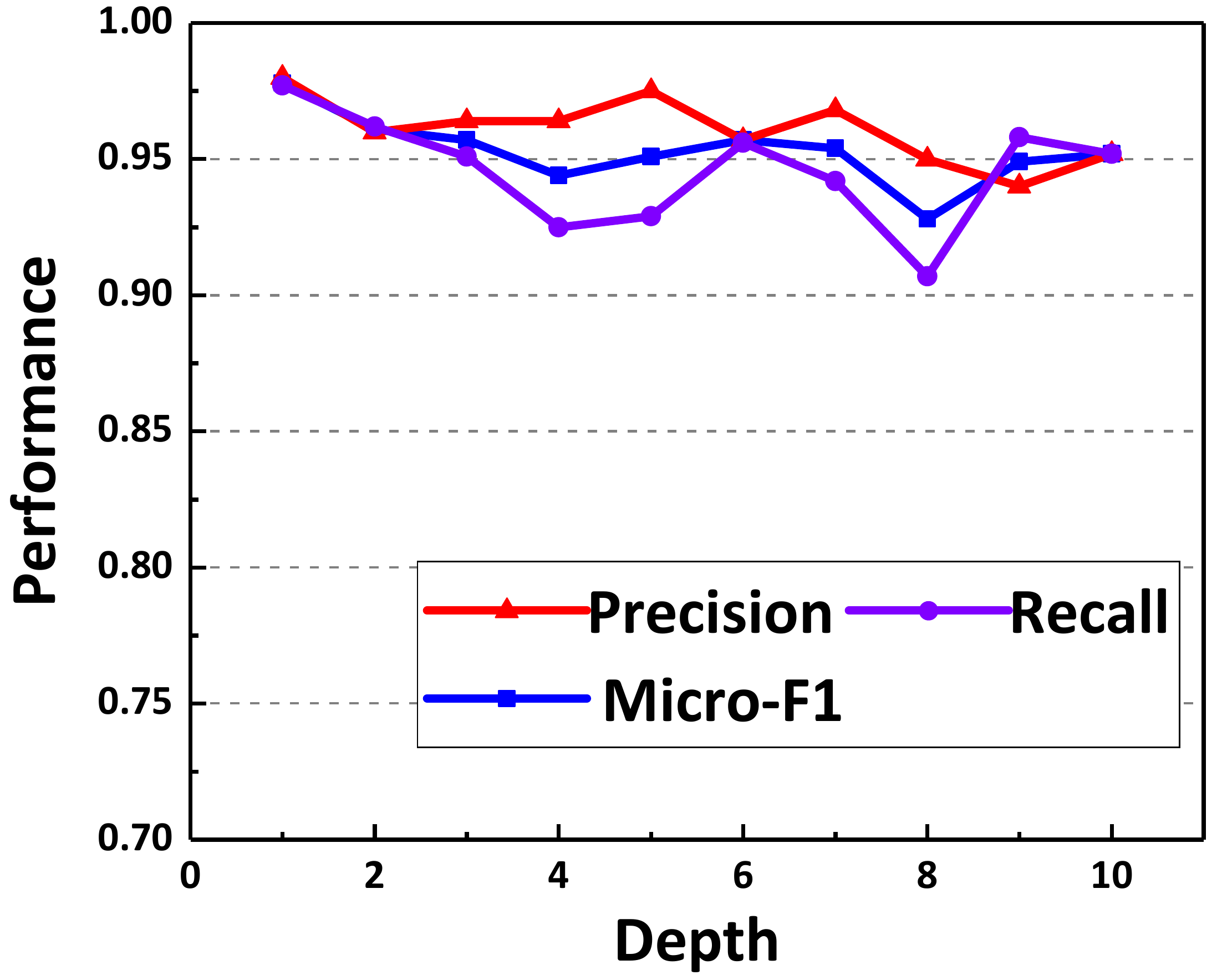}} 
  \subfigure{ 
    \includegraphics[width=1.3in]{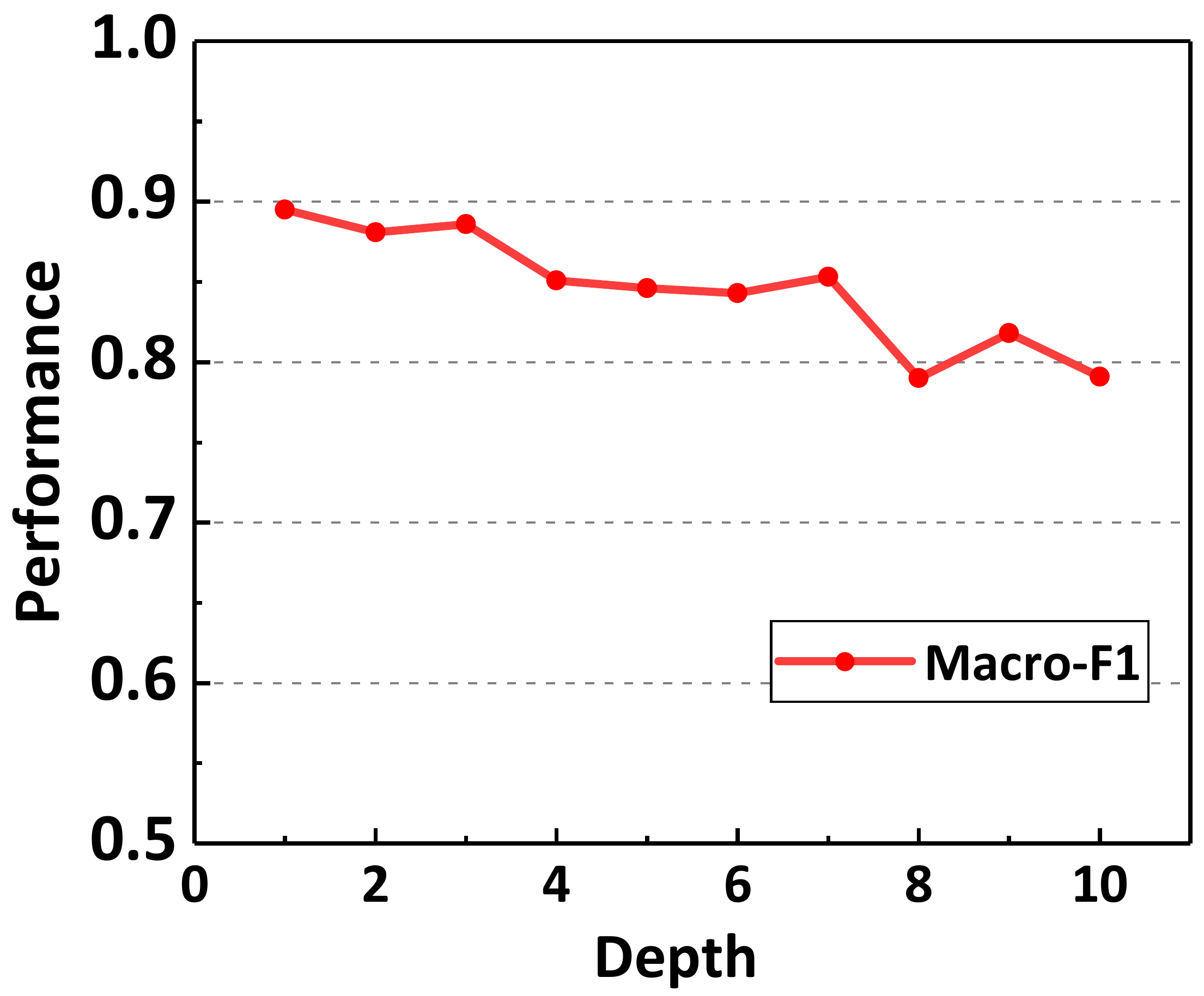}}
    
  \caption{Performance comparison on WIRCNN regarding word embedding size and convolutional depth, respectively.} 
  \label{Figure6}
\end{figure*}
%
%

%
%

\begin{figure*}
  \centering 
  \subfigure{
    \includegraphics[width=1.34in]{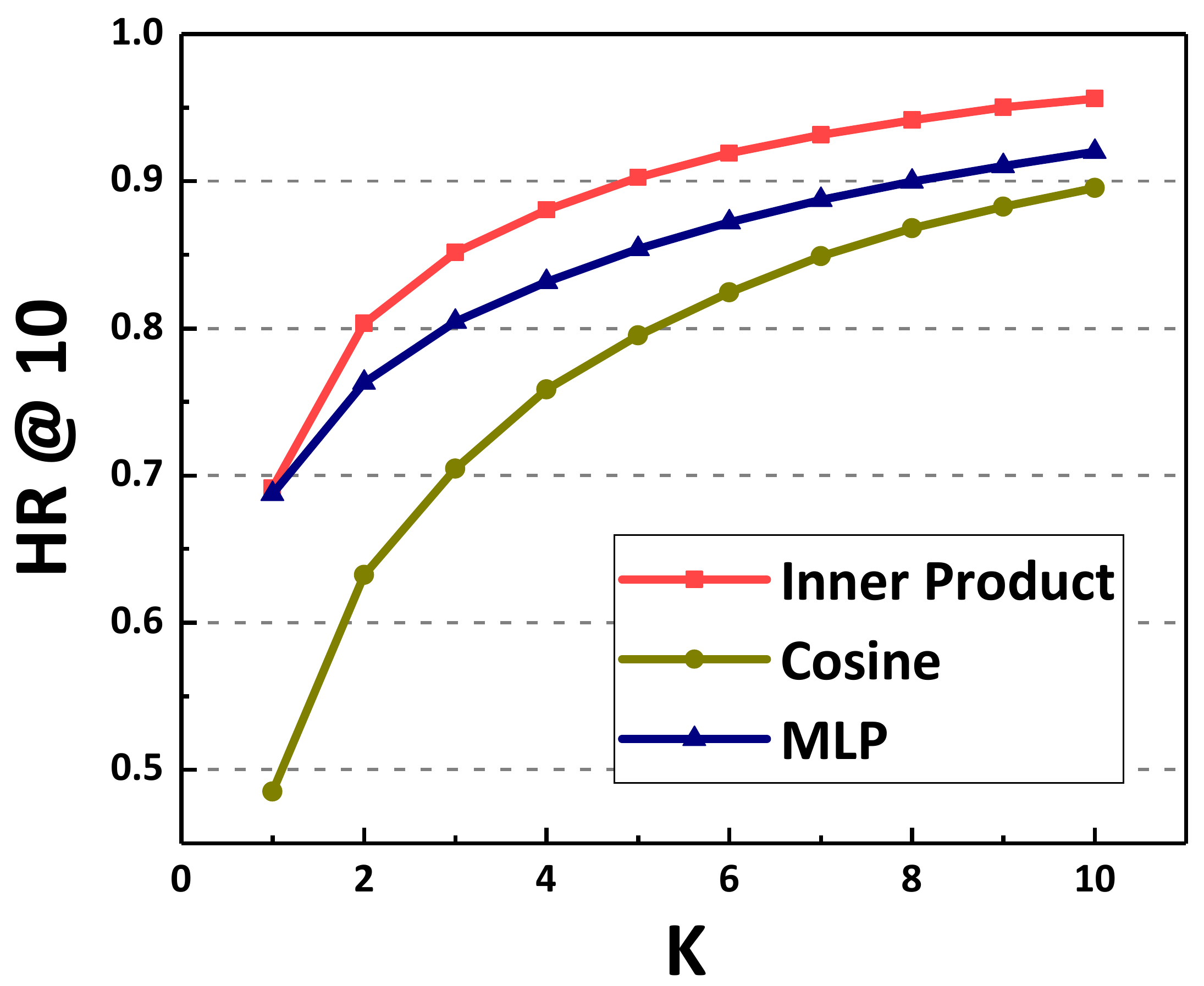}} 
  \subfigure{ 
    \includegraphics[width=1.34in]{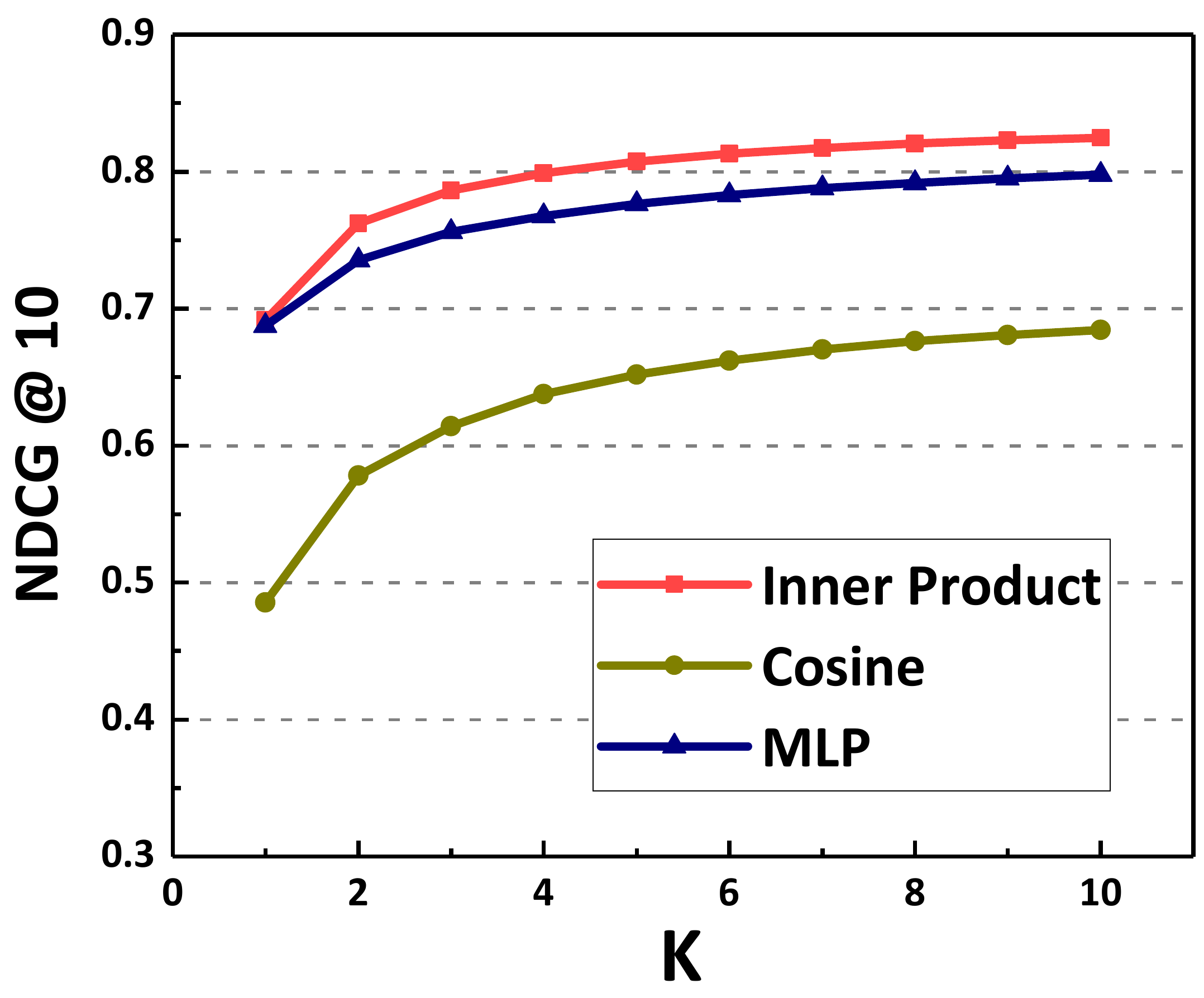}} 
      \subfigure{
    \includegraphics[width=1.27in]{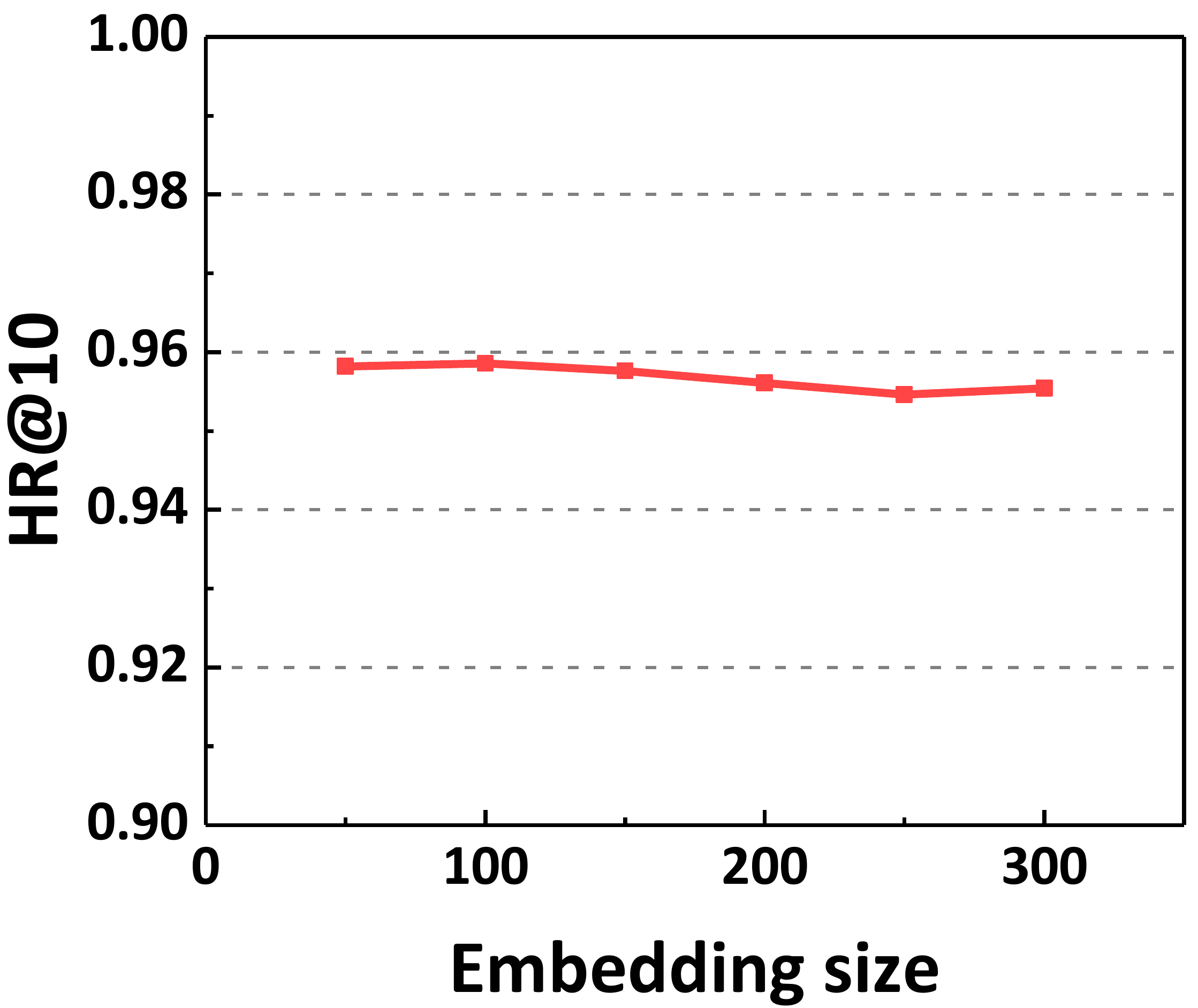}} 
  \subfigure{ 
    \includegraphics[width=1.27in]{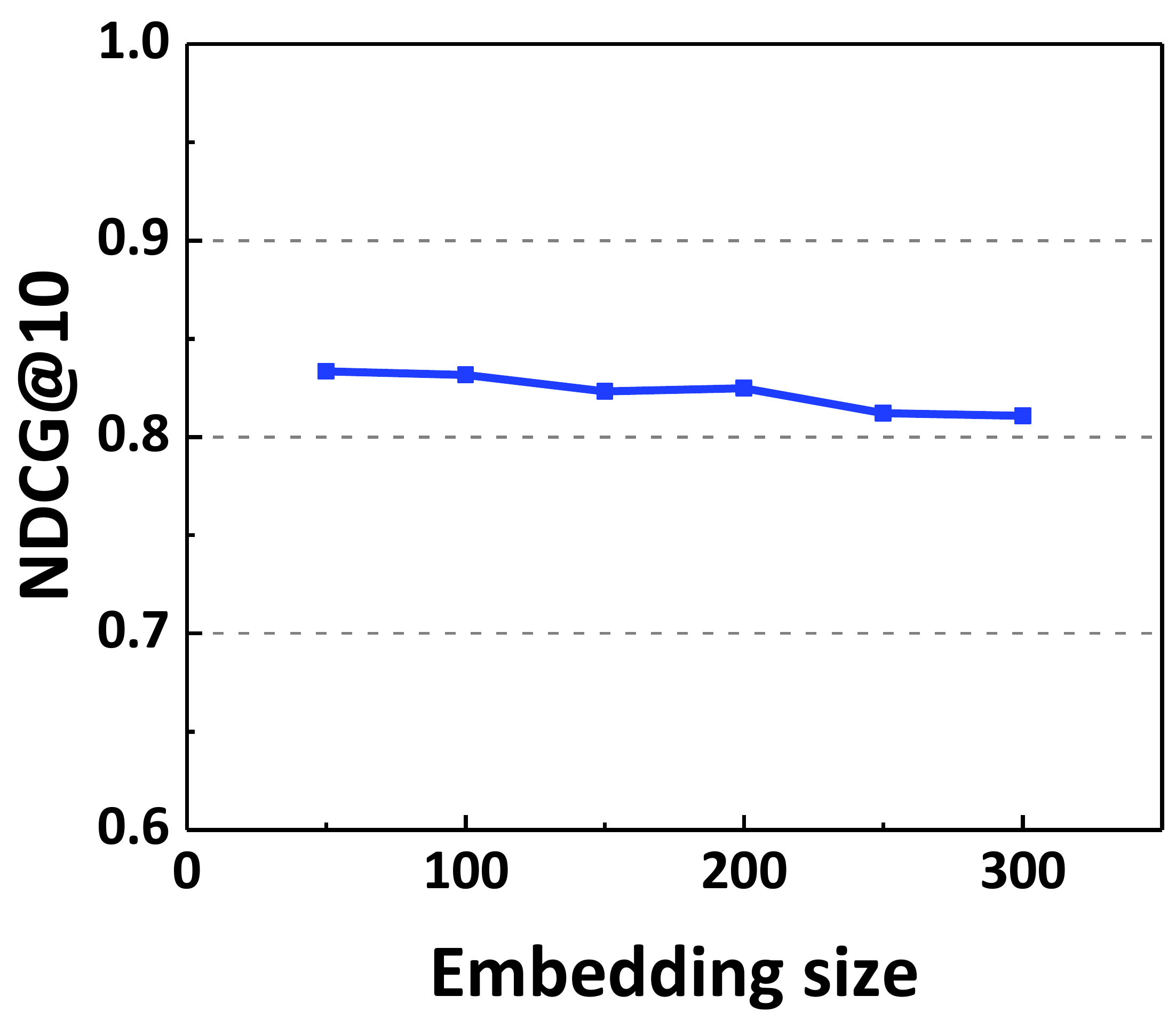}} 
  \caption{Evaluation of the proposed recommender with respect to the similarity function and the embedding size of users/recipes.} 
  \label{Figure7}
\end{figure*}
%
%
%
%
\subsubsection{Parameter Sensitivity}

To estimate the proposed models' sensitivity to hyper-parameters, we conducted many contrast experiments to measure the performance of WIRCNN and the recommender under different hyper-parameter settings. Note that only one hyper-parameter was changed at a time. The results of WIRCNN and the recommender are reported in Figure \ref{Figure6} and Figure \ref{Figure7}, respectively. Regarding WIRCNN, we evaluated it with different sizes of word embeddings and depths of convolutional layers. From Figure \ref{Figure6}, we can have the following observations: 1) at the beginning, the performance of WIRCNN is improving with the increase of the word embedding size, but it remains stable when the embedding size exceeds 50. This indicates that WIRCNN isn't very sensitive to the change of word embedding size; 2) the performance of WIRCNN drops slightly when the depth of convolutional layer increases, which proves that increasing the complexity of the neural networks blindly doesn't work well in this task. As to the health-aware recommender, we testified the effect of the similarity functions and the user/recipe embedding size. From the results shown in Figure \ref{Figure7}, we find that: 1) inner product is the best function to calculate the similarity score in this task and consine performs the worst; 2) the performance maintains the same level with the increase of embedding size. The proposed recommender is relatively insensitive to the embedding size of users and recipes.

%
%
%
\section{Conclusion and Future Work}
In this work, we present a personalized health-aware food recommendation scheme, consisting of three main components, namely recipe retrieval, user health profiling, and health-aware food recommendation. To justify our proposed deep models, we constructed two high-quality datasets. And experimental results demonstrate the effectiveness of our health-aware food recommendation scheme and the superior performance of the proposed models. Moreover, from the experimental results, we could draw the following conclusions: 1) attentive feature extraction is crucial to the user health profiling based on sparse data. And 2) the category-level information significantly matters in the food recommendation.

This work is only a small step to build the personalized health-aware food recommendation system. In the future, we will continue to perfect the proposed scheme from the following directions: 1) building more mature systems to profile the user health from multiple aspects. 2) Health-aware ingredient recommendation is also a promising research direction to protect the user health. And 3) incorporating more healthy diet knowledge in more effective ways to make recommendation. The embedding and incorporation of knowledge are emerging research topics whereas the existing work about the food-related knowledge is limited.



\begin{acks}
This work is supported by the National Key Research and Development Project of New Generation Artificial Intelligence, No.:2018AAA0102502; the National Natural Science Foundation of China, No.:61772310, and No.:U1936203; the Shandong Provincial Natural Science Foundation, No.:ZR2019JQ23; the Innovation Teams in Colleges and Universities in Jinan, No.:2018GXRC014; the Shandong Provincial Key Research and Development Program, No.:2019JZZY010118.
\end{acks}

\tiny
\bibliographystyle{ACM-Reference-Format}
\bibliography{bibfile}

\end{document}